\documentclass[smallextended,final]{svjour3}

\journalname{GRG}
\date{\today}
\usepackage{xspace} 
\usepackage[textsize=small]{todonotes}

\usepackage[color]{showkeys}
\definecolor{refkey}{rgb}{0.8,0.0,0.0}
\definecolor{labelkey}{rgb}{0.2,0.2,0.9}
\usepackage{amsmath,amsfonts}
\usepackage{mathtools}
\usepackage[scr=rsfso]{mathalfa}
\usepackage[utf8]{inputenc}
\usepackage{enumitem}

\newcommand*{\cF}{\mathcal{F}}

\newcommand*{\cO}{{\mathcal{O}}}

\newcommand*{\RR}{\mathbb{R}}

\newcommand*{\jb}{\mathbf{j}}
\newcommand*{\xb}{\mathbf{x}}
\newcommand*{\qb}{\mathbf{q}}
\newcommand*{\vb}{\mathbf{v}}
\newcommand*{\nb}{\mathbf{n}}
\newcommand*{\ab}{\mathbf{a}}
\newcommand*{\eb}{\mathbf{e}}
\newcommand*{\Vb}{\mathbf{V}}

\newcommand*{\sigmab}{{\boldsymbol{\sigma}}}
\newcommand*{\omegab}{{\boldsymbol{\omega}}}
\newcommand*{\gammab}{{\boldsymbol{\gamma}}}

\newcommand*{\sM}{\mathscr{M}}

\newcommand*{\ZZ}{\mathbb{Z}}
\newcommand*{\dd}{\mathrm{d}}
\newcommand*{\del}{\partial}
\newcommand*{\eps}{\epsilon}

\newcommand*{\ee}{{\mathrm{e}}}
\newcommand*{\Lie}{{\mathscr{L}}}

\title{Notes on the Sagnac effect in General Relativity}
\author{Jörg Frauendiener}
\institute{Department of Mathematics and Statistics, University of Otago, New Zealand \and
  Institut de Mathématiques de Bourgogne, Université de Bourgogne, Dijon, France\\
 \email{joerg.frauendiener@otago.ac.nz}}

\begin{document}
\maketitle
\begin{abstract}
  The Sagnac effect can be described as the difference in travel time between two photons traveling along the same path in opposite directions. In this paper we explore the consequences of this characterisation in the context of General Relativity. We derive a general expression for this time difference in an arbitrary space-time for arbitrary paths. In general, this formula is not very useful since it involves solving a differential equation along the path. However, we also present special cases where a closed form expression for the time difference can be given. The main part of the paper deals with the discussion of the effect in a small neighbourhood of an arbitrarily moving observer in their arbitrarily rotating reference frame. We also discuss the special case of stationary space-times and point out the relationship between the Sagnac effect and Fizeau's ``aether-drag'' experiment.
\end{abstract}

\section{Introduction}
\label{sec:introduction}

In 1913, Georges Sagnac published a short communication~\cite{Sagnac:1913tx} in the proceedings of the French Academy of Sciences describing an experiment to prove the existence of the ether. He elaborated on this a couple of months later~\cite{Sagnac:1913tz}. The experiment consisted of an interferometer on a rotating table in which two light rays propagating in opposite directions along the same path are brought to interference and a phase shift is detected depending on the angular velocity of the table and the area enclosed by the traveling light. We refer to this as the ``classical Sagnac effect''.

Unbeknownst to Sagnac, Franz Harre\ss, a German graduate student, conducted a similar experiment in 1911~(see the report by Knopf~\cite{Knopf:1920df}) where he considered counter-propagating light in a ring of totally reflecting prisms. His objective was entirely different from Sagnac and, in fact, his experiment did not agree with his expectations since he neglected the very effect that Sagnac exhibited. Max von Laue, in 1920 compared both experiments from a special relativistic point of view~\cite{vonLaue:1920fd}. 

The Sagnac effect has many applications and ramifications in experiments and technology. Post~\cite{Post:1967hu} gives an extensive review until 1967. Since then many new developments have occurred. We mention only a few which have relevance for Relativity. The Sagnac effect is present in the GPS and other systems of satellites and has to be accounted for in order to achieve the high precision of operation~\cite{Allan:1985gr}. In a similar way, the Hafele-Keating experiment~\cite{Hafele:1972gn} can be regarded as a manifestation of the Sagnac effect. The Sagnac effect is also present in counter-propagating matter waves, see~\cite{Hasselbach:1993kq}.

The Sagnac interferometer with laser light propagating in opposite directions \emph{along the same path} has several advantages over the Michelson interferometer as already recognised by Michelson himself. The classical Sagnac effect is proportional to the area enclosed by the path. However, this is the \emph{signed} area (for details see below). This implies that by choosing the path appropriately, one can make the classical Sagnac effect disappear. The resulting \emph{zero-area Sagnac} interferometers are insensitive to rotations (but not to accelerations, see below). Due to this property this type of interferometer has become of interest to the gravitational wave community. They are considered as a possible alternative to the traditional Michelson layout for third generation gravitational wave detectors, see~\cite{Huttner:2016ep,Bond:2017vy}

There are several theoretical improvements over the classical result. Ori and Avron~\cite{Ori:2016bf} present a special-relativistic analysis of deformable interferometers such as the Sagnac interferometer and give also an explanation of the Wang experiment~\cite{Wang:2004ef}. Tartaglia~\cite{Tartaglia:1998dn} derives general relativistic corrections to the classical Sagnac effect in a Kerr space-time.

In this paper we present an approach to discuss the Sagnac and related experiments in rather general situations. We rederive the classical effect and some corrections to it due to the motion of the reference frame defined by the laboratory and possible curvature effects. The plan of the paper is as follows. We explain our setup in sect.~\ref{sec:gener-sagn-effect} and derive the Sagnac effect in full generality. Sect.~\ref{sec:sagnac-effect-fermi} to \ref{sec:some-example-paths} are devoted to a discussion of three contributions to the effect which arise in a certain approximation based on (generalized) Fermi coordinates. In sect.~\ref{sec:stat-space-times} we apply our framework to general relativistic stationary space-times and sect.~\ref{sec:fizeau-experiment} discusses the Fizeau experiment as a special case.

The conventions used here are those of Penrose and Rindler~\cite{Penrose:1984wm}.

\section{The general Sagnac effect}
\label{sec:gener-sagn-effect}

Generally speaking, the Sagnac effect can be described as the difference in travel time between two photons traveling along the same path in opposite directions. In this section we will derive a formula for this quantity in general and then evaluate it in the following section using some reasonable approximations.

We assume that things happen in a 4-dimensional space-time $(\sM,g)$ with $g$ a Lorentzian metric. We choose an observer, i.e., a time-like world-line $O$ parameterized by proper time $\tau$ not necessarily a geodesic. Let $a^b$ be the acceleration of the world-line then the 4-velocity $t^a$ of $O$ satisfies the equation
\begin{equation}
  \label{eq:1}
  \nabla_t t^b = a^b.
\end{equation}
This motivates the introduction of the Fermi derivative $\cF_t$ along $O$ defined by
\begin{equation}
  \label{eq:2}
  \cF_t X^b = \nabla _t X^b + a_c X^c t^b - t_c X^c a^b
\end{equation}
for any vector field $X^a$ defined along $O$. The Fermi derivative has the well-known properties that \textbf{(i)} $t^b$ is Fermi constant along $O$: $\cF_t t^b = 0$ and \textbf{(ii)} it is compatible with the metric: $\cF_t g = 0$. The expression~\eqref{eq:2} is not the most general one satisfying the conditions \textbf{(i)} and \textbf{(ii)}. In fact, we can incorporate a skew-symmetric tensor $\omega_{ab} = -\omega_{ba}$ perpendicular to $t^b$ resulting in the generalized Fermi transport law
\begin{equation}
  \label{eq:3}
  \cF_t X^b - \omega^b{}_c X^c = 0
\end{equation}
We can now define a Fermi frame adapted to the observer $O$ by transporting a tetrad $(\eb_0,\eb_1,\eb_2,\eb_3)$ along $O$ using the generalized Fermi transport
\begin{equation}
  \label{eq:4}
  \cF_t \eb^b_k - \omega^b{}_c \eb^c_k = 0, \quad \text{for } k=0:3 \qquad  \text{and } \eb_0^a = t^a.
\end{equation}
Physically, this models an observer together with his lab in which he carries out experiments. The three space-like frame vectors span the lab, the space $\Sigma_\tau$ of simultaneity at a given instant of time $\tau$. The lab is allowed to accelerate and to rotate. When $a^b=0$ and $\omega^a{}_b=0$ then the reference frame is freely falling and non-rotating, i.e., it is an inertial frame.

The simultaneity spaces foliate $\sM$ near the world-line $O$ of the observer. For a given $\tau$ the space $\Sigma_\tau$ intersects $O$ perpendicularly at the proper time~$\tau$. We assume that this foliation is global and define a global time function $\tau: \sM \to \RR$ so that $\Sigma_\tau = \{\tau = \text{const}\}$. Then we can write the metric in the familiar $1+3$-form
\begin{equation}
  \label{eq:5}
  g = g_{00} \dd \tau^2 + 2 g_{0k} \dd \tau \dd x^k + g_{ik} \dd x^i \dd x^k
\end{equation}
where $(x^k)_{k=1:3}$ are arbitrary coordinates on $\Sigma_\tau$.

Let us now fix an arbitrary event on $O$ which we may label without loss of generality by $\tau=0$ and let $\gamma$ be an arbitrary closed curve in $\Sigma_0$ starting at $P\in\Sigma_0$ having the coordinates $x^i_0$. Let $v$ be the tangent vector to $\gamma$. Then we have $v^a\nabla_a \tau = 0$, i.e., $v = v^i\del_i$. In order to compute the travel time of a photon around $\gamma$ we need to find the null curve $\hat\gamma$ in $\sM$ which projects to $\gamma$. The null tangent vector $l$ to $\hat\gamma$ can be written as $l =  L\del_\tau + v^i\del_i$ with $L>0$. At each point of $\hat\gamma$ the equation
\begin{equation}
  \label{eq:6}
  g_{00} L^2 + 2 g_{0i}L v^i + g_{ik}v^iv^k = 0
\end{equation}
holds. It can be solved for $L$
\begin{equation}
  \label{eq:7}
  L = \frac1{g_{00}} \left(\sqrt{(g_{0i}g_{0k} - g_{00}g_{ik})v^iv^k} - g_{0k}v^k \right)
\end{equation}
where the sign was chosen so that $L>0$. We abbreviate this expression as $L(g,v)$.

The null curve $\hat\gamma(s) = (\tau(s),\xb(s))$, where $\xb(s)$ is an abbreviation for a parametrization $(x^i(s))_{i=1:3}$ of $\gamma$, can now be obtained by solving the system of differential equations
\begin{equation}
\left.  \begin{aligned}
  \dot \tau(s) &= L(\tau(s),\xb(s)) , \\ \dot x^k(s) &= v^k(\xb(s))
\end{aligned}\right\}
 \qquad \text{with } \tau(0) = 0, \quad \xb(0) = \xb_0.\label{eq:8}
\end{equation}
Since the solution of the spatial part of this equation is, by construction, the curve $\gamma$ we may regard $\xb(s)$ and hence $\vb(s) = \dot\xb(s)$ as known in the time component of \eqref{eq:8}. Then, this equation is of the form $\dot\tau(s) = L(g(\tau(s),\xb(s)),\vb(s)) = f(\tau(s),s)$ which cannot be solved explicitly unless we know $f$, i.e., the curve $\gamma(s)$ and the metric components.

For the sake of simplicity we take a parametrization of $\gamma$ with $s\in [0,1]$ so that $P=\gamma(0) = \gamma(1)$. Let $\tau(s)$ be the solution of \eqref{eq:8}. Then the travel time $T[\gamma]$ of the photon traveling along $\gamma$ is given formally by the integral
\begin{equation}
  \label{eq:9}
  T[\gamma] = \tau(1) = \int_0^1 L(g(\tau(s),\xb(s)),\vb(s))\, \dd s.
\end{equation}

Next, let $\tilde\gamma$ be the reversed curve parameterized by $\tilde\gamma(s) = \gamma(1-s)$ and let $\tilde\tau(s)$ be the solution of the equation
\begin{equation}
\dot{\tilde\tau}(s) =  L(g(\tilde\tau(s),\tilde\xb(s)),\tilde\vb(s)), \quad \tilde\tau(0) = 0.\label{eq:10}
\end{equation}
Then we obtain the travel time of a photon traveling in the opposite direction as
\[
  T[\tilde\gamma] = \tilde\tau(1) = \int_0^1 L(g(\tilde\tau(s),\tilde\xb(s)),\tilde\vb(s))\, \dd s.  
\]
Thus, the time difference $\Delta_\gamma T = T[\gamma] - T[\tilde\gamma]$ is $\Delta T = \tau(1) - \tilde\tau(1)$. In order to evaluate this formula for a given curve $\gamma$ in a space-time with a given metric $g$ we need to solve the differential equations \eqref{eq:8} and \eqref{eq:10}. 

We can rewrite this formula in a slightly different way. Using the relationship between the two paths we get $\tilde\xb(s) = \xb(1-s)$ and $\tilde\vb(s) = - \vb(1-s)$ and hence
\[
  L(g(\tilde\tau(s),\tilde\xb(s)),\tilde\vb(s)) =   L(g(\tilde\tau(s),\xb(1-s)),-\vb(1-s)) .
\]
and the travel time becomes
\begin{equation}
  \label{eq:11}
  T[\tilde\gamma] = \int_0^1 L(g(\tilde\tau(1-s),\xb(s)),-\vb(s))\, \dd s.  
\end{equation}
Thus, the time difference $\Delta T$ is given as one integral
\begin{equation}
  \label{eq:12}
  \Delta_\gamma T = \int_0^1 L(g(\tau(s),\xb(s)),\vb(s))  - L(g(\tilde\tau(1-s),\xb(s)),-\vb(s))\,\dd s.
\end{equation}

\section{The Sagnac effect in Fermi coordinates}
\label{sec:sagnac-effect-fermi}

The general expression~(\ref{eq:12}) for the time difference due to the Sagnac effect is implicit and too general to allow any detailed statements. One could start to approximate the solutions $\tau$ and $\tilde\tau$ by iterating the differential equations but this would lead to complicated formulae and probably to the same results as what we will do next.

To proceed further we introduce Fermi coordinates adapted to the observer world-line $O$. Then the metric coefficients up to terms cubic in $x$ are (see app.~\ref{sec:fermi-coordinates})
\begin{align}
      g_{00} &= 1 - 2 a_l x^l + 3 (a_mx^m)^2 + \omega_{im} \omega^i{}_n x^m x^n + R_{m0n0} x^m x^n + O(x^3),\label{eq:13}\\
    g_{0k} &= \omega_{kl} x^l  + \frac23  R_{m0nk} x^m x^n+ O(x^3),\label{eq:14}\\
    g_{kl} &= -\delta_{kl} + \frac13 R_{mlnk} x^m x^n+ O(x^3).\label{eq:15}
\end{align}
Here, $\omega_{ik}$, $a_k$ and the Riemann coefficients are functions of $\tau$ defined on $O$.

In order to proceed we will now make the following two reasonable but significant approximations:
\begin{itemize}[wide]
\item The expected time difference $\Delta T$ is small compared to the time scale of changes in the reference frame and the gravitational field.
\item The path $\gamma$ is not ``too extended'' so that we remain in the $O(x^3)$ neighbourhood of the coordinate system.
\end{itemize}
By the first assumption we can take $a^k$, $\omega_{ik}$ and the Riemann tensor coefficients as time independent. Then the metric coefficients are time independent so that $L(g(\tau,\xb),\vb) = L(g(\xb),\vb)$. Inserting this into the formula for $\Delta_\gamma T$ yields
\[
  \begin{aligned}
    \Delta_\gamma T &= \int_0^1 L(g(\xb(s)),\vb(s)) - L(g(\xb(s)),-\vb(s))\,\dd s \\
    &= -2 \int_0^1 \frac{g_{0i}}{g_{00}}v^i\, \dd s
\end{aligned}
  \]
  since the terms involving the square root cancel each other. We may express this formula in terms of the \emph{Sagnac 1-form} $\sigmab = -2\frac{g_{0i}}{g_{00}}\dd x^i$ as
  \begin{equation}
    \label{eq:16}
    \Delta_\gamma T = \int_\gamma \sigmab = \int_{S} \dd \sigmab
  \end{equation}
where $S$ is any spanning 2-surface bounded by $\gamma$\footnote{Such surfaces always exists. One particular class consists of minimal surfaces which can be constructed as solutions of Plateau's problem~\cite{Douglas:1931bf}}.

By the second assumption, the metric is given in the $\cO(x^3)$ neighbourhood defined by Fermi coordinates by the equations~(\ref{eq:13}-\ref{eq:15}). Up to cubic terms we find
\[
  \frac1{g_{00}} = 1 + 2 a_l x^l + (a_mx^m)^2  - \omega_{im} \omega^i{}_n x^m x^n - R_{m0n0} x^m x^n + O(x^3)
\]
and the Sagnac form becomes
\begin{equation}
  -2\frac{g_{0l}}{g_{00}} = -2\omega_{lk}x^k + 4 (a_mx^m) \omega_{lk}x^k - \frac43 R_{m0nl}x^m x^n + O(x^3).\label{eq:17}
\end{equation}
These are three terms with different characteristic properties. We discuss them in order in the next section.

\section{Discussion of the contributions}
\label{sec:discussion-effects}

In this section we will make use of the usual 3-vector notation writing $a^k$, $x^k$ as $\ab$, $\xb$ etc. This is justified by the fact that the vector field $x^i\del_i$ behaves like the Euclidean position vector $\xb$ to the order we are interested in. We also introduce the angular velocity vector $\omegab$ with components $\omega^i = \frac12\eps^{ikl}\omega_{kl}$ and the area form $\dd^2S$ of a surface $S$ by $\dd x^l\wedge\dd x^k = \eps^{lkm}n_m\,\dd^2S$, where $n_k$ is the unit-normal to $S$. Also, we note that we raise and lower spatial indices $i,j,k,\ldots$ with $\delta_{ik} = -\eta_{ik}$. This allows us to also make use of the notation $\xb\cdot \ab = x_i a^i$ for the usual Euclidean inner product.

\subsection{The pure rotation effect}
\label{sec:pure-rotation-effect}

Consider the first term in \eqref{eq:17}. It depends only on the angular velocity and contributes a time difference of
\[
  \Delta_\omega T = -2\int_\gamma\omega_{lk}x^k \dd x^l = 2\int_S \omega_{lk}\dd x^l \wedge \dd x^k = 4 \int_S (\omegab\cdot \nb)\, \dd^2S.
\]
Thus, the time difference due to this term is a multiple of the ``rotation flux'' through the surface $S$ spanned by $\gamma$. This contribution is invariant under translations~\cite{Schwartz:2017jx}, i.e., under the transformation $x^k \mapsto x^k + q^k$ for constant $q^k$ and rotations $x^k \mapsto \alpha^k{}_l x^l$, where $\alpha^k{}_l$ is a constant orthogonal matrix.

This is the classical Sagnac effect as described in~\cite{Sagnac:1913tz} for, if we assume that $\omega_{ik}$ describes a rotation around the $3$-axis, i.e., when $\omegab = \omega \eb_3$ and that, further, $\gamma$ is a simple closed curve in the $(12)$-plane then we obtain
\[
  \Delta_\omega T = 4 \omega \int_S (\nb\cdot \eb_3)\,\dd^2S = \pm4 \omega\, \text{area}(S).
\]
Of course, the sign depends on the relative orientation between the angular velocity and the curve~$\gamma$.

It is easy to design curves for which the time difference vanishes. Trivial examples are curves which lie in a plane parallel to the rotation axis. Non-trivial curves have the shape of a figure eight or something more complicated when projected parallel to the rotation axis. These curves give rise to the ``zero-area'' Sagnac configurations~\cite{Bond:2017vy}. More complicated examples are straightforward to construct.

\subsection{The acceleration dependent effect}
\label{sec:accel-depend-effect}
Next, we consider the second term in~\eqref{eq:17}. This term, which vanishes when $a^k=0$, contributes a time difference
\[
  \Delta_a T = 4  \int_\gamma (a_mx^m) \omega_{lk}x^k \dd x^l = -8 \int_S  a_{(k} \omega_{m)l} x^m \dd x^k\dd x^l .
\]
Again, this can be written as the flux of a vector field $\Vb$ through the surface $S$ spanned by $\gamma$. In this case, the vector field is $\Vb = (\ab \cdot \omegab) \xb - 3  (\ab \cdot \xb) \omegab$ so that
\begin{equation}
  \label{eq:18}
  \Delta_a T = 4  \int_S (\ab \cdot \omegab) (\nb\cdot \xb) - 3  (\ab \cdot \xb) (\omegab\cdot\nb)\,\dd^2S.
\end{equation}
In contrast to the pure rotation effect, the acceleration effect is \emph{not} translation invariant: under a translation $\xb\mapsto \xb + \qb$ the time difference changes according to
\[
  \Delta_a T \mapsto \Delta_a T  + 4  \int_S (\ab \cdot \omegab) (\nb\cdot \qb) - 3  (\ab \cdot \qb) (\omegab\cdot\nb)\,\dd^2S.
\]
This means that we can make the time difference vanish by shifting the curve.

\subsection{The gravitational effect}
\label{sec:gravitational-effect}

Finally, we come to the third term which is entirely due to the gravitational field in the form of the Riemann tensor. Before the detailed discussion we need to briefly digress to introduce the decomposition of the Riemann tensor $R_{abc}{}^d$ into the Schouten or ``Rho'' tensor $P_{ab}$ (representing the Ricci tensor~\cite{Penrose:1984wm}) and the Weyl tensor $C_{abc}{}^d$.
\[
  R_{ab}{}^{cd} =   C_{ab}{}^{cd} -4 \delta_{[a}{}^{[c} P_{b]}{}^{d]}.  
\]
The right dual of the Riemann tensor is
\[
  R^\star_{abcd} = C^\star_{abcd} - \eps_{abc}{}^{e} P_{ed} + \eps_{abd}{}^e  P_{ce} .
\]
With $t^a$ the time-like 4-velocity of the observer we obtain
\[
  t^ct^d  R^\star_{acbd}  = t^ct^dC^\star_{acbd} - t^c t^d \eps_{acb}{}^{e} P_{ed}.
\]
With the relationship $P_{ab} =  -\frac12 \left(G_{ab} - 12 g_{ab} G\right)$ and the Einstein equation $G_{ab} = - 8\pi T_{ab}$ this becomes
\[
 t^ct^d  R^\star_{acbd}  = B_{ab} - 4\pi t^dt^c\eps_{abc}{}^{e}  \left(T_{de} - \frac16 g_{de} T\right) =  B_{ab} - 4\pi \eps_{ab}{}^{c}j_c.
\]
Here, we have used the definitions $\eps_{abc} = t^e\eps_{eabc}$, $t^eT_{ea} = j_a$ and $B_{ab} = C^\star_{acbd}t^ct^d$ for the 3-dimensional volume form, the momentum density and the magnetic part of the Weyl tensor with respect to the time-direction $t^a$. Expressed in terms of the Fermi coordinates this equation becomes\footnote{keeping in mind that indices $i$, $k$ etc.\ are moved with $\delta_{ik} = -\eta_{ik}$.}
\begin{equation}
  \label{eq:19}
 R^\star_{i0k0}  = B_{ik} + 4\pi \eps_{ik}{}^{l}j_l.
\end{equation}
The time delay from the Riemann tensor is
\[
  \Delta_R T = \frac43 \int_\gamma R_{0mnl}x^m x^n \,\dd x^l = 4 \int_S R_{0(mn)l}x^n\dd x^m\wedge\dd x^l.
\]
It can be expressed in terms of the right dual of the Riemann tensor
\begin{equation}
  \label{eq:20}
\Delta_R T = 4 \int_S R^\star_{0n}{}^{0i}x^nn_i\dd^2S
\end{equation}
which in turn, using~\eqref{eq:19}, can be cast into the form
\begin{equation}
  \label{eq:21}
  \Delta_R T = 4 \int_S B_{i}{}^k x^in_k\dd^2S - 16\pi \int_S (\xb \times \jb)\cdot \nb\,\dd^2S.
\end{equation}
This shows that the gravitational Sagnac effect in the first order is entirely due to ``magnetic'' interaction, both in the gravitational wave part due to $B_{ab}$ and the matter part due to the  flux of angular momentum density through $S$. As the acceleration effect, the gravitational effect is also not translation invariant.

\section{Some example paths}
\label{sec:some-example-paths}

In order to get some idea about how different shapes of paths influence the time delay we now consider a restricted class of paths, 3-dimensional Lissajous curves, given in parameterised form by
\begin{equation}
  \label{eq:22}
  \gammab(s) = 
  \begin{bmatrix} 
    A_1 \sin (l s + \alpha_1)\\
    A_2 \sin (m s + \alpha_2)\\
    A_3 \sin (n s + \alpha_3)
   \end{bmatrix},
   \qquad s,\alpha_1,\alpha_2,\alpha_3 \in [0, 2\pi], \quad l,m,n \in \ZZ.
\end{equation}
We assume these curves have period $2\pi$ which implies $\gcd(l,m,n) = 1$, i.e., $l$, $m$ and $n$ are relatively prime. It is straightforward to insert the parameterisation into the expressions for the time delay. Again, we discuss the different contributions sequentially.

\subsection{The rotation term}
\label{sec:rotation-term}

By choosing the axes of the frame appropriately we can arrange that $\omegab = \omega\,\eb_3$ and then the time delay due to $\omega$ becomes
\begin{equation}
  \label{eq:23}
    \Delta_\omega T =  4\pi m\omega\,A_1 A_2 \left(\delta_{l,-m}\sin(\alpha_1+\alpha_2) + \delta_{l,m} \sin(\alpha_1-\alpha_2) \right).
\end{equation}
This shows, that $\Delta T$ vanishes unless the projection of the curve perpendicular to $\omegab$ is a non-degenerate ellipse. Choosing $l\ne \pm m$ yields ``zero-area'' paths. Since the behaviour of the curve in the direction of $\omega$ is irrelevant these paths can be chosen without self-intersections.

\subsection{The acceleration term}
\label{sec:acceleration-term}

Keeping $\omegab$ along the $\eb_3$ axis we can rotate the frame around $\omegab$ to make the acceleration vector $\ab$ lie in the plane spanned by $\eb_1$ and $\eb_3$. Then we can write $\ab = a(\eb_3 + \lambda \eb_1)$ for some real $\alpha$ and $\lambda$. With these simplifications we can write the contribution of the acceleration term due to the class of curves~\eqref{eq:23} as $\Delta_\ab T = 8\pi A_1A_2 a \,\omega J(l,m,n)$ where  
\begin{equation}
  \label{eq:24}
 \begin{aligned}
J(l,m,n) &=   A_1 \lambda \left[(l-m) \cos (2
  \alpha_1+\alpha_2)\,\delta_{m,-2l} - (l+m) \cos (2 \alpha_1-\alpha_2)\,\delta_{m,2l} \right] \\
+ &A_3 \left[(l+m) \bigl(\cos (\alpha_1-\alpha_2-\alpha_3)\, \delta_{n,l-m} - \cos(\alpha_1-\alpha_2+\alpha_3)\,\delta_{n,m-l}\bigr)\right.\\
  +& \left.(l-m) \bigl(\cos (\alpha_1+\alpha_2+\alpha_3) \delta_{n,-(l+m)} - \cos (\alpha_1+\alpha_2-\alpha_3)\,\delta_{n,l+m}\bigr)\right].
  \end{aligned}
\end{equation}
The first two terms in this expression are due to the misalignment of angular velocity vector and acceleration. They vanish for $\lambda =0$. Let us first discuss this case. Then the time delay is proportional to $A_1A_2A_3$, i.e., to the volume of the rectangular box which contains the space curve. It vanishes unless at least one of the four equations
\begin{equation}
  \label{eq:25}
  n+l+m = 0,\quad
  n+l-m = 0,\quad
  n-l+m = 0,\quad
  n-l-m = 0
\end{equation}
holds. It is easy to see that for a non-degenerate curve these equations cannot hold simultaneously. It is also not possible for just one of them to be violated. Thus, at most two of the equations can hold simultaneously. When two equations hold then it follows that one integer must vanish, while the other two are equal in magnitude and then they must be equal to $\pm1$. In these cases, the curve is planar, being contained in a plane perpendicular to one of the coordinate axes. If this plane is perpendicular to the $\eb_i$-axis, then it is a distance $A_i\sin(\alpha_i)$ away from the origin. Thus, one can make the time-delay vanish by choosing $A_i=0$ or $\alpha_i=0$. This is a consequence of the translation dependence of the acceleration term.

Due to the geometry, the case $n=0$ is different from $l=0$ which, in turn, is equivalent to $m=0$. In the former case, we obtain for $l=m=1$ (the case $l=-m$ can be obtained by reversing the orientation of the curve and replacing $\alpha_2$ by its negative) 
\[
J(1,1,0) = 4 A_3 \sin(\alpha_1-\alpha_2)\sin(\alpha_3)
\]
while the case $l=0$ with $n=m=1$ yields
\[
J(0,1,1) = 2 A_3 \sin(\alpha_1-\alpha_3)\sin(\alpha_2),
\]
the case $m=-1$ again corresponding to an orientation reversal.

For the general case, when only one of equations~\eqref{eq:25} holds, we may take as an example $n=l+m$. Then $l$ and $m$ are non-zero and relatively prime and the corresponding curve is non-planar. Its contribution becomes
\[
J(l,m,l+m) = -A_3(l-m) \cos(\alpha_1+\alpha_2-\alpha_3),
\]
which is non-zero unless the phases are chosen in a very specific way.

When angular velocity and acceleration are not aligned then there are two additional possible terms in~\eqref{eq:24}. They are proportional to $\lambda$ and they contribute only when $m=\pm2l$. This condition does not involve $n$ which can therefore be chosen so that one of the equations~\eqref{eq:25} is satisfied. One possibility is $l=1$, $m=2$, $n=3$ which yields
\begin{equation}
  \label{eq:26}
  J(1,2,3) = -3 A_1 \lambda \cos(2\alpha_1-\alpha_2) + A_3 \cos(\alpha_1+\alpha_2-\alpha_3).
\end{equation}

\subsection{The gravitational term}
\label{sec:gravitational-term}

As mentioned in sec.~\ref{sec:gravitational-effect} this term contains two contributions, one due to the Weyl tensor and another due to the matter. We first discuss the Weyl term. It is mediated by the magnetic part $B_{ik}$ of the Weyl tensor. Let us assume that this term is due to a gravitational wave propagating in the $\eb_3$ direction. Then $B_{ik}$ has the form
\[
   B_{ik} =
  \begin{bmatrix}
    a_1 & a_2 & 0 \\
    a_2 & -a_1 & 0 \\
    0 & 0 & 0 
  \end{bmatrix}
\]
for some real constants $a_1$ and $a_2$. Inserting the parameterisation for the curves~\eqref{eq:22} we find the time delay for the Weyl contribution to be
\begin{equation}
  \label{eq:27}
  \begin{aligned}
\Delta_B T =  
\frac{3\pi}2 a_1 A_1 A_2 A_3 n \;\bigl( &\cos(\alpha_1 - \alpha_2 - \alpha_3)\;\delta_{n,l-m}
+ \cos(\alpha_1 - \alpha_2 + \alpha_3)\;\delta_{n,m-l}\\
- &\cos(\alpha_1 + \alpha_2 - \alpha_3)\;\delta_{n,l+m}
- \cos(\alpha_1 + \alpha_2 + \alpha_3)\;\delta_{n,-l-m}\bigr)\\
+\frac{3\pi}2 A_3 a_2\bigl(
l A_1^2  &\left[\cos(2 \alpha_1 - \alpha_3)\,\delta_{n,2l} - \cos(2 \alpha_1 + \alpha_3)\,\delta_{n,-2l} \right]\\ +  m A_2^2 &\left[\cos(2 \alpha_2 + \alpha_3)\,\delta_{n,-2m} - \cos(2 \alpha_2 - \alpha_3)\delta_{n,2m}\right]\bigr).
\end{aligned}
\end{equation}
This comes in two pieces each corresponding to a different polarisation state of the wave. The first is proportional to $a_1$ and corresponds to the $+$-polarisation. Its 'signature' is the same as the one for the aligned acceleration case --- they are non-zero for the same class of curves. As an example we pick a curve with $n=l+m$ and obtain
\begin{equation}
  \label{eq:28}
\Delta_+ T =  
-\frac{3\pi}2 a_1 A_1 A_2 A_3 n \; 
 \cos(\alpha_1 + \alpha_2 - \alpha_3)
\end{equation}
The $\times$-polarisation contributes the term proportional to $a_2$. It has the same signature as the misaligned acceleration case. It is non-zero only if the curve has a figure eight projection in a direction perpendicular to the propagation of the wave. Choosing $n=2l\ne2|m|$ yields the contribution
\begin{equation}
  \label{eq:29}
\Delta_\times T =  
\frac{3\pi}2 A_1^2 A_3 a_2
l  \cos(2 \alpha_1 - \alpha_3).
\end{equation}

\section{Stationary space-times}
\label{sec:stat-space-times}

As a further application we discuss the Sagnac formula~\eqref{eq:12} in a stationary space-time $\sM$ where we have a time-like Killing vector $\xi^a$. The length of the Killing vector is a scalar function on $\sM$ defined by
\begin{equation}
  \label{eq:30}
  \xi_a\xi^a = \ee^{2U} 
\end{equation}
and we have the following relations
\begin{equation}
  \label{eq:31}
  \Lie_\xi g_{ab} = -2\nabla_{(a}\xi_{b)} = 0, \quad \xi^a\nabla_aU = 0, \quad \nabla_a\xi_b = 2 \nabla_{[a}U \xi_{b]} + \omega_{ab},
\end{equation}
where $\omega_{ab} = -\omega_{ba}$ and $\xi^a\omega_a = 0$.

We write $t^a:=\ee^{-U}\xi^a$ for the unit-vector in the direction of $\xi^a$. We also pick one integral curve $O$ of $\xi^a$. Since $U$ is constant along $O$ we can scale $\xi^a$ to become a unit-vector along $O$. With $t$ the parameter along $\xi^a$, i.e., $\xi^a\nabla_at=1$ we find that $t$ measures proper time for an observer along $O$, i.e., with 4-velocity $t^a = \xi^a$. The metric can be written in the form
\[
g = g_{00} \dd t^2 + 2 g_{0k}\dd t \dd x^k + g_{ik} \dd x^i \dd x^k
\]
with $\del_tg_{\mu\nu}=0$ and $g_{00}=\ee^{2U}$. With transformations of the form $t\mapsto t+\alpha_ix^i$ for constants $\alpha^i$ we can arrange that $g_{i0}=0$ on $O$ and $x^i \mapsto x^i + \beta^i{}_kx^k$ for constants $\beta^i{}_k$ achieves that $g_{ik} = - \delta_{ik}$ on $O$.
In terms of these coordinates, $\xi^a \doteq \del_t$ and $\xi_a \doteq g_{0\mu}\dd x^\mu$. 

Furthermore, the observer along $O$ is accelerated since 
\[
a_b = t^a\nabla_at_b = \ee^{-2U} \xi^a\nabla_a \xi_b = -\nabla_b U.
\]
We now set up a \emph{stationary frame} for the observer on $O$ by choosing in a neighbourhood of $O$ three  vector fields $\eb_i$ which, together with $t^a$, form an orthonormal basis along $O$ and which are invariant under $\xi^a$, i.e., for which $\Lie_\xi \eb_i=0$ holds. These vector fields can be chosen to be the coordinate vector fields $\del_i$.

We are now in the same situation as for the derivation of the Sagnac formula~\eqref{eq:12}. Due to the stationarity of the space-time there is no dependence on $t$ and we can evaluate the integrals as before resulting in the same formula
\begin{equation}
  \label{eq:32}
  \Delta T = -2 \int_\gamma \frac{g_{0i}}{g_{00}}\,\dd x^i,
\end{equation}
except that now this formula is exact. The integrand is easily identified as the pull-back to the curve of the $1$-form $\alpha_a:=\xi_a/(\xi_c\xi^c)$, the ``inverted Killing vector''. Using the Stokes theorem as before we can write the integral as a surface integral over a spanning surface $S$ for the curve $\gamma$ of the $2$-form
\[
  \nabla_{[a}\alpha_{b]} = \frac{\nabla_{[a}\xi_{b]}}{\xi_c\xi^c} - \frac{\xi_{[b}\nabla_{a]} \left( \xi_c\xi^c\right)}{(\xi_c\xi^c)^2} = \ee^{-2U} \left( \nabla_{[a}\xi_{b]} - 2\xi_{[b}\nabla_{a]} U\right) = \ee^{-2U} \omega_{ab}.
\]
The time difference therefore becomes
\begin{equation}
  \label{eq:33}
  \Delta T = -2 \int_S \ee^{-2U}\omega_{ik}\dd x^i\wedge\dd x^k.
\end{equation}
The quantity $\omega_{ik}$ is in fact closely related to the angular velocity of the stationary frame with respect to a locally non-rotating Fermi transported frame. This can be seen by comparing Fermi- and Killing transport: let $v^a$ be Lie dragged along the Killing vector so that $\xi^c\nabla_cv^a = v^c\nabla_c \xi^a$ holds. We compute the Fermi derivative of $v^a$ along the unit-vector $t^c$
\[
  \begin{multlined}
  \cF_t v^a = t^c\nabla_c v^a + t^aa_cv^c - a^at_c v^c = \ee^{-U}\left( v^c\nabla_c \xi^a + \xi^aa_cv^c - a^a\xi_c v^c\right) \\= \ee^{-U}\left( v^c(\xi^a \nabla_cU - \xi_c \nabla^aU + \omega_c{}^a )+ \xi^aa_cv^c - a^a\xi_c v^c\right) = \ee^{-U}\left( v^c\omega_c{}^a\right)
\end{multlined}
\]
which shows that the angular velocity of the stationary frame with respect to the Fermi frame is $-\ee^{-U}\omega_{ik}$. In contrast to the discussion in sect.~\ref{sec:gener-sagn-effect}, here the formula is exact. The acceleration terms which appear there correspond to the factor $\ee^{-2U}$ here. This factor partly corrects for the difference between the Killing time $t$ and proper time and partly serves to introduce the ``gravitational force'' $\nabla_aU$ which is responsible for the acceleration.

\section{Light in a moving medium and the Fizeau experiment}
\label{sec:fizeau-experiment}

As a final example we consider Minkowski space $\mathbb{M}$ with its flat metric $\eta_{ab}$ filled with a homogeneous, isotropic medium which is transparent and without dispersion. Then the light rays move with a different velocity $\bar{c}$ which is related to the speed of light in vacuum $c$ by $\bar{c} = c/n$ where $n$ is the refraction index of the material, defined in terms of its permittivity $\eps$ and permeability $\mu$. By assumption these,  and therefore $n$, are constant. The material is described by a 4-velocity $u^a$ with $u_au^a = 1$.

Let $t^a$ be the 4-velocity of an observer. To simplify things we assume that $t^a$ is covariantly constant. Together with a spatial frame of covariantly constant unit-vectors $t^a$ forms a basis and we can introduce global Cartesian coordinates $(t,x^i)$ on $\mathbb{M}$.

Following \cite{Gordon:1923vm} and \cite{Ehlers:1967dk} we describe the motion of the light by null geodesics with respect to the ``optical metric''
\begin{equation}
  g_{ab} = \eta_{ab} - u_a u_b (1-1/n^2).\label{eq:34}
\end{equation}
The observer splits the matter 4-velocity $u^a$ into time and space components\footnote{We use the Minkowski metric $\eta_{ab}$ for moving indices.},
\[
  u^a = \gamma \left(t^a + v^a\right), \qquad\text{with } \gamma = (1-v^2)^{-\tfrac12}, \quad t_av^a=0
\]
where we have defined $v^2 := -v_av^a = \mathbf{v}\cdot\mathbf{v}$.

The 4-velocity $t^a$ is a time-like Killing vector for $\eta_{ab}$ and, assuming that $u^a$ is Lie dragged along $t^a$, the optical metric has the property that $\Lie_t g_{ab}=0$. Thus, we are in the situation of sect.~\ref{sec:stat-space-times} describing a stationary system.

Considering a spatial path $\gamma$ traversed by light in opposite directions we find in general the time difference given by~\eqref{eq:32}. Evaluating the metric coefficients we find
\[
  g_{00} = 1-(1-1/n^2) \gamma^2, \qquad g_{0i} = -(1-1/n^2)\gamma^2 v_i.
\]
If the material velocity is such that its spatial part $v^a$ has closed stream-lines then we can take the path $\gamma$ of the light to be parallel to a stream-line of length $L$, traversing it parallel to the motion of the medium. Thus, we may write $v^i = v \dot x^i$ if we assume parametrisation of $\gamma$ by arc-length. Finally, let us assume that $v$ is constant along a stream-line then we obtain for the time difference for the light moving along $\gamma$ in opposite directions
\[
  \Delta T = 2 \int_\gamma \frac{(1-1/n^2)\gamma^2}{ 1-(1-1/n^2) \gamma^2} v_i\dot{x}{}^i(s) \, \dd s = - 2 \frac{(1-1/n^2)\gamma^2}{ 1-(1-1/n^2) \gamma^2} v L.
\]
This formula can be simplified to
\begin{equation}
  \label{eq:35}
  \Delta T = -2L v \frac{n^2 - 1}{1 - n^2 v^2}.
\end{equation}

This setup describes the classical ``aether-drag'' experiment by Fizeau\cite{Fizeau:1851ta} to determine the difference of the speed of light in a medium moving in opposite directions. For a very nice summary of that experiment and the classical theoretical background we refer to the paper by Lahaye et. al.~\cite{Lahaye:2012fa}. The classical derivation makes use of the special relativistic addition formula for velocities to find the speed of light (in units of $c$) in (opposite to) the direction of a moving medium (water) as
\[
  v_{\pm} = \frac{v \pm 1/n}{1\pm v/n}.
\]
Therefore, the difference in travel time along a path of length $L$ in opposite directions is
\[
  \Delta T = \frac{L}{v_+} - \frac{L}{v_-} = -2L v \frac{n^2 - 1}{1 - n^2 v^2}
\]
in complete agreement with~\eqref{eq:34}. This shows that the effects of Sagnac and Fizeau are merely facets of the same coin, a fact which seems to have been suspected for some time, see e.g.,~\cite{Leeb:1979wa}.

Obviously, the formula~\eqref{eq:35} given above, can be readily generalised to non-homogeneous media and cases where the light path is not aligned with a closed stream line. However, it may not be possible to give a closed form expression.

\section{Conclusion}
\label{sec:conclusion}

In this paper we derived the Sagnac effect, i.e., the difference in travel time for light moving in opposite directions along the same spatial path, from first principles within Einstein's general theory of relativity. The resulting formula is difficult to evaluate in full generality since one needs to solve a differential equation along the path. We have considered several special cases where the evaluation is possible. 

The first case addressed a general space-time but considered only a small neighbourhood around an observer. Introducing Fermi coordinates and assuming that the travel time of the light is much smaller than the time scale for changes in the observers frame we were able to give a closed expression for the time difference. Within the approximation used, there are three contributions to this time difference: the first one is the classical Sagnac effect caused by the rotation of the reference frame. The next order term is caused by a combination of the rotation and the acceleration of the frame and, in the same order, there is a contribution from the curvature of the space-time. The structure of the terms is such that one can set up (combinations of) light paths which are sensitive to one single term only. One possible application of this could be to measure acceleration and rotation of a reference frame or, alternatively, to measure gravitational wave signals. Having said that, one should point out that we have not at all discussed the size of the expected time differences in concrete situations.

We should also point out that our approach can not immediately take care of experiments in the spirit of Wang~\cite{Wang:2004ef} since we assume that the path is contained in a hyper-surface of constant time. In contrast, in the Wang setup the path is allowed to change its shape during the experiment. However, it should be possible to take care of this effect in a more or less straightforward way within our framework.

The second case we discussed involved stationary space-times. We found that the time difference there was caused by the flux of the curl of the ``inverted Killing vector'', which amounts to the rescaled rotation part of the Killing vector. In this case, the time difference is due to the appropriately measured angular velocity of the stationary frame with respect to a freely falling frame. Again, without looking at the size of the effect, this could provide a means to measure the dragging of inertial frames in a rotating gravitational system.

Finally, we specialised the stationary case to a moving homogeneous and isotropic medium. We showed that with the use of the appropriate optical metric it is possible to reproduce the classical explanation for the Fizeau experiment, which demonstrated the dependence of the speed of light on the relative motion between medium and observer.

\section{Acknowledgments}
\label{sec:acknowledgment}

I wish to thank the CNRS of France for a visiting position at the Département de Mathématiques at  Université de Bourgogne in Dijon, France where some of this research was carried out. My thanks also go to Eyal Schwartz for sparking my interest in the Sagnac effect and to Niels Kjaergaard for pointing me to reference~\cite{Lahaye:2012fa}.
\appendix

\section{Fermi coordinates}
\label{sec:fermi-coordinates}

Fermi coordinates cover a neighbourhood of an observer, i.e., a time-like line $O$ in the space-time $\sM$. They are adapted to the observer in the sense that the metric when expressed in these coordinates assumes the Minkowski form $\mathrm{diag}(1,-1,-1,-1)$ on the world-line. Fermi coordinates can be obtained by first defining a tetrad along the world-line by Fermi-Walker transport and then using the exponential map at each point of $O$ restricted to the subspace perpendicular to $O$ to assign coordinates to an entire neighbourhood of the world-line.

Since the exponential map is in general only a local diffeomorphism the coordinate chart is not global. Usually, the Fermi coordinates are derived only for inertial observers, see the classic paper by Manasse and Misner~\cite{Manasse:1963ha} for a very lucid treatment. Synge~\cite{Synge:1976vf} describes the Fermi coordinates for an accelerated observer carrying along a non-rotating frame. This is not enough for our purposes since we are interested also in rotating reference frames. Therefore, we have to repeat the derivation as given in~\cite{Manasse:1963ha} with the presence of acceleration and rotation in mind.

Since the calculations are not very illuminating and follow exactly along the same steps as described by Manasse and Misner we will not repeat them here. A short description of the process should suffice. We pick a time-like world-line $O$ with unit-tangent vector $t^a$. Then the parameter $\tau$ along the world-line measures proper time for the observer. The world-line need not be a geodesic so that it may have a non-zero acceleration $a^b = t^a\nabla_at^b$. Then, Fermi-Walker transport of a vector (field) $X^a$ along $O$ is defined by the equation
\begin{equation}
  \label{eq:36}
  \cF_t X^b := t^a\nabla_a X^b + (t^b a_c - a^bt_c)X^c = 0.
\end{equation}
This transport law corrects for the instantaneous boost that occurs due to the acceleration of the observer with respect to a locally freely falling observer. Two vector fields transported along $O$ in this way maintain their inner product. However, \eqref{eq:36} is not the most general transport law with that property. We are free to add a piece involving an infinitesimal rotation in the  3-space perpendicular to $t^a$. Thus, we are led to consider vectors $X^a$ transported according to
\begin{equation}
  \label{eq:37}
  \cF_t X^b + \omega^b{}_c X^c =0,
\end{equation}
where $\omega_{ab} = \omega_{ba}$ is defined along the world-line and satisfies $\omega_{ab}t^b=0$.

Now we consider three spatial unit-vectors $e_i^a$ with $i=1,2,3$ at a point on $O$ which are mutually orthonormal and perpendicular to $t^a$. We transport them along $O$ according to~\eqref{eq:37} with some $\omega_{ab}$ which we consider as given. Next, we consider families of geodesics $C(\lambda;\tau,x^a)$ starting at points with proper time $\tau$ on $O$ with initial tangent vector $x^a$ perpendicular to $t^a$. We can now assign coordinates to all points which can be reached after one unit of affine parameter $\lambda$: let $P = C(1;\tau,x^ie_i^a)$ then we assign to $P$ the coordinates $(\tau,x^i)$.

The next goal is to obtain an expression for the metric coefficients in the Fermi coordinate system. This can only be done in terms of an expansion in the coordinates $x^i$. The lowest order of the metric coefficients when $x^i=0$ is, by construction,
\[
  g_{00} = 1, \quad
  g_{0i}  = 0, \quad
  g_{ik}  = -\delta_{ik}.
\]
The next step is to compute the first derivatives of the metric coefficients on $O$ and this is done using the transport law for the basis and $t^a$ to find the Christoffel symbols on $O$. The result of this computation is
\[
    \Gamma^0_{00} = 0, \quad \Gamma^i_{00} = a^i, \quad \Gamma^0_{0k} = - a_k, \quad \Gamma^l_{0k} =  \omega^l{}_k, \quad \Gamma^l_{ik} = 0.
\]
Using the expression of the Christoffel symbols in terms of the first derivatives of the metric yields
\[
  \begin{aligned}
    \del_0 g_{00} &= 0\\
    \del_0 g_{0k} &= \Gamma_{00k} + \Gamma_{k00} = 0,\\
    \del_0 g_{ik} &= \Gamma_{i0k} + \Gamma_{k0i} = 0,\\
  \end{aligned}\qquad
  \begin{aligned}
    \del_l g_{00} &= \Gamma_{0l0} + \Gamma_{0l0} = -2a_l,\\
    \del_l g_{0k} &= \Gamma_{0lk} + \Gamma_{kl0} = \omega_{kl}, \\
    \del_l g_{ik} &= 0.
  \end{aligned}
\]
Finally, one needs to compute the second derivatives of the metric coefficients. This is done by evaluating the equation for geodesic deviation which involves the components of the Riemann tensor on $O$. We will not present the details of the calculation here but simply state the result ($\alpha$ and $\beta$ take values $0,\ldots,3$)
\[
  \begin{aligned}
    \del_0\del_0 g_{\alpha\beta} &= 0, \quad \del_0\del_l g_{00} = -2\dot a_l, \quad \del_0\del_l g_{0k} = \dot \omega_{kl}, \quad \del_0\del_l g_{ik} = 0,\\
    \del_m\del_n g_{kl} &= \frac13 \left( R_{mlnk} + R_{mknl}\right),\\
    \del_m\del_n g_{k0} &= \frac23 \left( R_{m0nk} + R_{n0mk}\right),\\
    \del_m\del_n g_{00} &= 6a_ma_n + 2\omega_{im} \omega^i{}_n + 2 R_{m0n0}.
  \end{aligned}
\]
Collecting the terms that we found in the various orders yields the metric in Fermi coordinates up to cubic terms in the spatial coordinates
\begin{align}
    g_{00} &= 1 - 2 a_l x^l + 3 (a_mx^m)^2 + \omega_{im} \omega^i{}_n x^m x^n + R_{m0n0} x^m x^n,\label{eq:38}\\
    g_{0k} &= \omega_{kl} x^l  + \frac23  R_{m0nk} x^m x^n,\label{eq:39}\\
    g_{kl} &= -\delta_{kl} + \frac13 R_{mlnk} x^m x^n.\label{eq:40}
  \end{align}

\providecommand{\bysame}{\leavevmode\hbox to3em{\hrulefill}\thinspace}
\providecommand{\MR}{\relax\ifhmode\unskip\space\fi MR }
\providecommand{\MRhref}[2]{%
  \href{http://www.ams.org/mathscinet-getitem?mr=#1}{#2}
}
\providecommand{\href}[2]{#2}

\end{document}